\documentclass{jpp}
\usepackage{graphicx}
\usepackage[T1]{fontenc}
\usepackage{amsmath}
\usepackage{bm}
\usepackage{natbib}
\usepackage{float}
\usepackage[colorlinks=true,
  linkcolor=black, citecolor=blue, urlcolor=blue]{hyperref}   % Modify colors of links
\usepackage[dvipsnames,svgnames,x11names,hyperref]{xcolor}  % Load colors for Tikz.
\usepackage[capitalize]{cleveref}

\usepackage{setspace}
\crefformat{subsection}{\S#1}

\shorttitle{Investigation of triangularity effects on edge turbulence gyrokinetic simulations}
\shortauthor{A C.D. Hoffmann et al.}

\title{Investigation of triangularity effects on tokamak edge turbulence through multi-fidelity gyrokinetic simulations}
\author{A.C.D. Hoffmann \aff{1}\corresp{\email{antoine.hoffmann@epfl.ch}}, P. Ricci \aff{1}}
\affiliation{\aff{1} Ecole Polytechnique F\'ed\'erale de Lausanne (EPFL), Swiss
Plasma Center, CH-1015 Lausanne, Switzerland
}
 \newcommand{\squareparenthesis}[1]{\left[#1\right]} 
 \newcommand{\roundparenthesis}[1]{\left(#1\right)} 

 \newcommand{\PJ}[2]{$(P,J)=(#1,#2)$}

 \newcommand{\RNa}{R_{Na}}
 \newcommand{\RTa}{R_{Ta}}

 \newcommand{\kperp}{k_\perp}
  
 \newcommand{\lperpa}{l_{\perp a}}

 \newcommand{\vpar}{v_\parallel}

 \newcommand{\Upsbar}{\bar \Upsilon}
 \newcommand{\Apar}{A_\parallel}

 \newcommand{\spara}{s_{\parallel a}}
 \newcommand{\wperpa}{w_{\perp a}}

 \newcommand{\dz}{\mathrm dz}
 \newcommand{\ddt}{\partial_t}

 \newcommand{\ddy}{\partial_y}
 \newcommand{\ddz}{\partial_z}

 \newcommand{\ddspara}{\partial_{s\parallel a}}

 \newcommand{\grad}{\nabla}

\newcommand{\kernel}{\mathcal{K}}

\newcommand{\ExB}{\bm E\times\bm B}

\newcommand{\gyacomo}{\textsc{Gyacomo} }

\usepackage{natbib}
\usepackage{graphicx}
\usepackage{bm}
\usepackage{cancel}
\usepackage{float}
\usepackage{algorithm}
\usepackage{algpseudocode}
\usepackage{multicol}

\begin{document}

\maketitle

\begin{abstract}
This paper uses the gyro-moment (GM) approach as a multi-fidelity tool to explore the effect of triangularity on tokamak edge turbulence. 
Considering experimental data from an L-mode DIII-D discharge, we conduct gyrokinetic (GK) simulations with realistic plasma edge geometry parameters at $\rho=0.95$. 
We find that employing ten GMs effectively captures essential features of both trapped electron mode (TEM) and ion temperature gradient (ITG) turbulence. 
By comparing electromagnetic GK simulations with adiabatic electron GK and reduced fluid simulations, we identify the range of validity of the reduced models.
We observe that TEMs drive turbulent heat transport under nominal discharge conditions, hindering accurate transport level estimates by both simplified models.
However, when TEMs are absent, and turbulence is ITG-driven, an agreement across the different models is observed. 
Finally, a parameter scan shows that the positive triangularity scenario destabilizes the TEM, therefore, the adiabatic electron model tends to show agreement with the electromagnetic simulations in zero and negative triangularity scenarios. 
On the other hand, the reduced fluid simulations exhibit limited sensitivity to triangularity changes, shedding light on the importance of retaining kinetic effects to accurately model the impact of triangularity turbulence in the tokamak edge. 
In conclusion, our multi-fidelity study suggests that a GM hierarchy with a limited number of moments is an ideal candidate for efficiently exploring triangularity effects on micro-scale turbulence.
\end{abstract}

%\ioptwocol
\section{Introduction}
% While being of significant importance in determining the performance of future fusion devices, turbulence in the tokamak plasma edge remains challenging to study.
% Computer simulations face formidable obstacles arising from the intricate, multi-scale nature of the phenomena taking place in the plasma edge, including various micro-instabilities -- both electromagnetic, such as kinetic ballooning modes (KBMs) and micro-tearing modes (MTMs), and electrostatic, such as ion temperature gradient (ITG), electron temperature gradient (ETG) and trapped electrons modes (TEMs). 

% % Positive experimental evidence
% Experimental findings from TCV \citep{Coda2022EnhancedTCV, Pochelon1999EnergyHeating}, DIII-D \citep{Austin2019AchievementTokamak} and ASDEX Upgrade \citep{Happel2023OverviewTokamak} demonstrate that negative triangularity (NT) magnetic equilibria exhibit confinement properties in the L-mode regime, comparable to the H-mode regime with positive triangularity (PT) \citep{Gohil1994, Wagner2007}.
% These confinement performances are achieved by avoiding a steep pressure gradient at the plasma edge, a condition known to trigger edge localized modes (ELMs) that have a detrimental impact on the vessel walls \citep{Zohm1996}.
% The potential for an ELM-free, high-confinement regime makes NT configurations a promising approach to the development of a long-lasting, economically viable fusion power plant like DEMO \citep{Marinoni2019H-modeDIII-D,Marinoni2021DivertedPedestal, Schwartz2022ToReactor}.

The theoretical understanding of the experimental observations that reveal confinement improvement in negative triangularity (NT) with respect to positive triangularity (PT) configurations remains an open question \citep{Coda2022EnhancedTCV, Pochelon1999EnergyHeating, Austin2019AchievementTokamak, Happel2023OverviewTokamak}.
Global plasma turbulence simulations, performed with fluid codes such as GBS \citep{Lim2023EffectConfigurations}, and gyrokinetic (GK) codes such as ORB5 \citep{Giannatale2022TriangularitySimulations} and GENE \citep{Merlo2021NonlocalPlasmas}, are able to reproduce the trends observed experimentally, notably the reduction of fluctuation level in NT.
However, the high computational cost of global simulations limits the possibility of carrying out large parameter scans, thus preventing, e.g., a thorough analysis of the mechanisms behind the improvement of plasma confinement in NT, depending on the plasma parameters, and the optimization of the design of future devices.

%(which is not the case for TOKAM3X \citep{Laribi2021ImpactSimulations}).
Despite the well-established role of global effects in high-performance tokamak scenarios \citep{Holland2011AdvancesPlasmas}, local GK simulations also provide evidence for a reduction of turbulent transport in NT discharges \citep{Happel2023OverviewTokamak, Ball2023LocalShear}.
% \cite{Duff2022EffectTurbulence} investigate the impact of NT on ITG-driven turbulence using local (i.e., flux tube) collisionless GENE simulations.
% Their findings reveal that the peak of the ITG growth rate occurs at the ion Larmor radius scale in the PT scenario.
% Considering NT, a shift toward larger poloidal wavelengths is observed, with a broader unstable wavenumber spectrum in the radial direction.
% Nonlinear simulations, carried out with adiabatic electrons, show a relatively weak dependence of the saturated heat flux level on triangularity, and comparisons with quasilinear results demonstrated a good agreement between the two approaches.
For example, the role of NT on the ion temperature gradient (ITG) instability is understood through collisionless GENE simulations considering an adiabatic electron model \citep{Duff2022EffectTurbulence}.
However, other instabilities may play a role in setting the level of edge turbulent transport, in addition to the ITG. 
In particular, trapped electron modes (TEMs) can be unstable in the edge of L-mode discharges, showing high sensitivity to the magnetic geometry \citep{Merlo2023OnTokamaks,Balestri2024PhysicalPlasmas}.
The possible role of electron-driven instability questions the adiabatic electrons assumption in studying the effect of triangularity on edge turbulence, requiring the use of costly multiscale GK simulations for the study of TEM-driven turbulence.

The cost of local GK simulations can be reduced by using the recently developed moment approach to the GK Boltzmann equation.
This is based on representing the velocity space dependence of the GK distribution function through an expansion on a Hermite-Laguerre polynomial basis, leading to the evolution of the moments of the distribution function that we denote as the gyro-moments (GMs) \citep{Jorge2017ACollisionality, Mandell2018,Frei2020}.
Linearly, \cite{Frei2023Moment-basedModel} show that ITG and TEM instabilities can be accurately evolved with the GM approach but a large number of GMs is necessary to resolve the sharp features of the passing-trapping boundary in the velocity space.
In nonlinear simulations, a large reduction of computational cost is observed in entropy mode and ITG scenarios when comparing the number of GMs with the number of points in a grid-based code required for convergence \citep{Hoffmann2023GyrokineticOperators,Hoffmann2023GyrokineticShift}.
However, it is unclear if the GM approach is also efficient to simulate TEM-driven turbulence.
% \cite{McClenaghan2023TransitionPlasmas} demonstrate a transition in the type of instability dominating the top of the pedestal in a DIII-D discharge, where the initially dominant ITG instability is superseded by a MTM instability when a critical value of the electron density is exceeded.
% Additionally, the impact of the adiabatic electron assumption remains an open question.
% \cite{Belli2023SpectralPedestal} present the first full multiscale simulations of pedestal-like transport, where both ion-scale and electron-scale fluctuations are simultaneously evolved.
% Their results indicate that the experimental parameters extracted from a PT H-mode DIII-D discharge reside in a bifurcation region, where ETG-driven transport becomes subdominant once a specific critical background temperature gradient is attained.

% Several key questions require further investigation to understand the effect of triangularity on tokamak edge turbulence.
% We mention three of them.
In the present paper, we perform linear and nonlinear GK flux-tube simulations of the edge region ($\rho=0.95$) of an L-mode DIII-D discharge  \citep{Boyes2023MHDPlasmas}. 
This discharge is part of a series of experiments where the magnetic shape is varied while keeping the plasma equilibrium profiles similar \citep{Boyes2023MHDPlasmas}.
%thus enhancing the realism of our theoretical scans in triangularity $\delta$.
Using the GM approach, we solve the GK Boltzmann equation in a flux tube, incorporating kinetic ions, kinetic electrons, and electromagnetic (EM) fluctuations.
The magnetic equilibrium is represented by using the Miller model \citep{Miller1998NoncircularModel}.
Confirming \cite{Hoffmann2023GyrokineticShift}, the GM approach properly describes GK turbulence with a reduced number of basis elements, enabling a significant reduction of the computational cost with respect to standard approaches.
To the authors' knowledge, the present nonlinear GK flux-tube simulations are the first ones to consider such an external flux surface in an experimental scenario, where the high magnetic shear requires a high spatial resolution.
%mentioning the efforts done in \cite{Hassan2021GyrokineticRegion}, where nonlinear simulations are briefly described, and in \cite{Neiser2019GyrokineticPlasmas}, where $\rho=0.90$ is considered. 
% Our only nonphysical assumption here resides in a halved ion-electron mass ratio ($ m_i/m_e = 10^3$), which allows an increase in the time-stepping.
\\
We first study the fastest-growing linear instability considering the experimental DIII-D equilibrium parameters. %and different values of triangularity $\delta$. 
The convergence of the linear growth rate with the number of GMs is then investigated.
Second, nonlinear simulations are performed to assess the capability of the GM model to resolve TEM-driven turbulence using a reduced GM basis for the first time.
Then, the multi-fidelity capabilities of the GM approach allow us to identify the mechanisms underlying the effect of triangularity through a comparison of a hierarchy of models. 
We compare nonlinear simulations obtained with the full GK model that includes kinetic electron (KEM), an adiabatic electron model (AEM), and a reduced fluid model (RFM) obtained from a hot electron asymptotic limit \citep{Ivanov2022DimitsTurbulence,Hoffmann2024thesis}.
For this comparison, we neglect the density gradient to favor ITG-driven turbulence, which the AEM and RFM simulations can also represent.
Our study sheds light on the multiple mechanisms responsible for the confinement enhancement observed in NT configurations and offers guidance for optimizing tokamak performance.
Moreover, it determines the range of validity of the simplified models and demonstrates that a ten-moment-based fluid system is a good candidate for efficiently studying triangularity effects, also when TEMs are present.

% In particular, we observe that reducing the triangularity can stabilize the TEM, stabilize the ITG but also simply reduce of the turbulence level.

This paper is organized as follows. 
The GM approach to the GK Boltzmann equation is briefly reviewed in Section \ref{sec:gkmodel} (details are provided in \ref{app:emgmhierarchy}).
Section \ref{sec:exp_prof_and_num} introduces the experimental DIII-D discharge employed as a reference and details the numerical setup of the modeling approaches we consider.
GK simulations of the DIII-D discharge at $\rho=0.95$ are presented in Sec. \ref{sec:nominal_parameters_simulations}, analyzing the convergence of the results with respect to the number of GMs. 
The results are also compared to AEM and RFM simulations.
The density gradient is then neglected and the triangularity is varied in Section \ref{sec:triangularity_scan}, to study and compare the predictions of the effect of triangularity on KEM, AEM and RFM simulations.
Section \ref{sec:conclusions} discusses the key findings and insights gained from investigations presented here.

\section{Gyrokinetic modelling based on the gyro-moment approach}
\label{sec:gkmodel}
We introduce the gyrokinetic distribution function for species $a$, $F_a(\bm r, v_\parallel, \mu,t)$, where $\bm r$ is the gyro-center position, $v_\parallel$ the component of the velocity parallel to the magnetic field $\bm B$, and $\mu$ the magnetic moment \citep{Catto1978LinearizedGyro-kinetics, Frieman1982NonlinearEquilibria}.
The distribution function is written as a sum of an equilibrium stationary Maxwellian background, $F_{a0}(\bm r, v_\parallel, \mu)$, and a fluctuating part, $g_a(\bm r, v_\parallel, \mu,t)$, i.e. $F_a=F_{a0} + g_a$.
We consider the $\delta f$ limit, $g_a/F_{a0}\sim \Delta \ll 1$ with $\Delta$ measuring the perturbation amplitude \citep{Hazeltine2003PlasmaConfinement}, and we evolve the distribution function in a flux tube domain considering the local limit.
Hence, we write the gyrocenter position using the field-aligned coordinates $\bm r = (x,y,z)$, where $\grad x$ is parallel to the minor radius of the torus, $\grad z$ is the unit vector parallel to the magnetic field and $\grad y$ is along the binormal direction.
Within these assumptions, the GK EM Boltzmann equation writes in dimensionless units \citep{Frei2023Moment-basedModel},
\begin{align}
    \frac{\partial g_a}{\partial t} &+ \{\Upsbar,g_a\}_{xy}+ \frac{\tau_a }{q_a}\squareparenthesis{\roundparenthesis{2 \spara^2+\wperpa} \mathcal C_{xy} - 2\spara^2 \alpha \ddy} h_a
    \nonumber\\&
    + \frac{1}{J_{xyz}\hat B}\frac{\sqrt{\tau_a}}{\sigma_a}\frac{\sqrt{2}}{2}\squareparenthesis{2\spara \ddz h_a - \wperpa\ddz\ln B\ddspara h_a} 
    \nonumber\\&
    +\squareparenthesis{R_{Na} + \roundparenthesis{s_\parallel^2+w_\perp-\frac{3}{2}} R_{Ta}}\ddy \Upsbar
    = \sum_b \nu_{ab} C_{ab}.
    \label{eq:3d_gyboeq_nondim}
\end{align}
In Eq. \ref{eq:3d_gyboeq_nondim}, $\Upsbar= \bar\phi-\vpar \bar\Apar$ denotes the gyro-averaged EM potential, where $\phi$ is the electrostatic potential and $\Apar$ the parallel component of the magnetic potential vector (the fluctuations of the magnetic field parallel to its equilibrium direction are neglected).  
We also introduce the non-adiabatic part of the distribution function, $h_a=g_a + F_{a0}\frac{q_a}{\tau_a}\Upsbar$, as well as the collision operator between species $a$ and $b$, $C_{ab}$, and the magnetic curvature operator, $\mathcal C_{xy}$. 
We use the analytical Miller geometry model to express $\mathcal C_{xy}$, the normalized magnetic field amplitude $\hat B$, and the Jacobian of the field-aligned coordinate system $J_{xyz}$ \citep{Miller1998NoncircularModel}, which depend on the aspect ratio, $\varepsilon$, safety factor, $q_0$, and shear, $\hat s$, of the considered flux surface. 
In addition, the shape of the flux surface is parameterized through the elongation, $\kappa$, triangularity $\delta$, squareness, $\zeta$, and their derivatives, $\hat s_\kappa$, $\hat s_\delta$ and $\hat s_\zeta$, respectively.
The normalized parameters and variables appearing in Eq. \ref{eq:3d_gyboeq_nondim} are presented in Tab. \ref{tab:dimensionless_units}.
\begin{table}
    \centering
    \begin{tabular}{l r l| l r l}
    Parallel velocity     & $\spara$ & $= v_{\parallel a}^{ph}/v_{th a}$  
    &%\\
    Perpendicular velocity     & $\wperpa$ & $= \mu^{ph}_a B_0/T_a$ 
    \\
    Wave numbers & $k_{x,y}$ & $= k_{x,y}^{ph}\rho_{s}$
    &%\\
    Normalized time & $t$ & $= t^{ph} c_{s}/R_0$
    \\
    Density gradient & $R_{Ta}$ & $= R_0/L_{Ta}$ 
    &%\\
    Temperature gradient & $R_{Na}$ & $= R_0/L_{Na}$ 
    \\
    Electric charge & $q_a$ & $= q_a^{ph}/e$ 
    &%\\
    Temperature ratio & $\tau_a$ & $= T_a/T_e$
    \\
    Particle mass ratio & $\sigma_a^2$ & $= m_a/m_e$ 
    &%\\
    Distribution function & $g_a$ & $= g^{ph}_a/F_{aM}$ 
    \\
    EM potential & $\Upsilon$ & $= e\Upsilon^{ph}/T_{e} $
    &
    Collision frequency & $\nu_{ab}$ & $= \nu^{ph}_{ab}/\nu_{ii}$
    \end{tabular}
    \caption{Dimensionless variables used in the GM model. 
    For a dimensionless variable $A$, its equivalent in physical units is explicitly denoted as $A^{ph}$.}
    \label{tab:dimensionless_units}
\end{table}

The GK Poisson equation for the EM potential, assuming the Debye length to be much smaller than the Larmor radius, yields the quasi-neutrality equation
\begin{equation}
	\sum_a \frac{q_a^2}{\tau_a}\roundparenthesis{1- \Gamma_0(\lperpa)}\phi = \sum_a q_a \int \mathrm d \bm v J_0 g_a,
	\label{ch2_eq:poissongk_units}
\end{equation}
where $\Gamma_0(\lperpa) = I_0(\lperpa) e^{-\lperpa}$, with $I_0$ the zeroth-order modified Bessel function of the first kind, and $\lperpa=\tau_a \sigma_a^2 \kperp^2 /2$ with $\kperp$ the perpendicular wavenumber.
% The right-hand side of Eq. \ref{ch2_eq:poissongk_units} represents the polarisation due to the guiding-center transform of the distribution function. 
% The left-hand side is the gyro-center density projected in the particle coordinates.
% It is noteworthy that the Boussinesq approximation, where the electric permittivity is considered constant, emerges naturally from the flux tube limit of the Poisson equation \citep{Held2023BeyondApproximation}.
Similarly, the Ampère equation yields a relation for the parallel magnetic potential,
\begin{equation}
	\left(2\kperp^2 + \beta_e \sum_a \frac{q_a^2}{\sigma_a^2}\Gamma_0(\lperpa)\right)A_\parallel = \beta_e \sum_a \frac{q_a \sqrt{\tau}}{\sigma_a}  \int \mathrm d \bm v J_0\vpar g_a.
	\label{ch2_eq:Ampèregk_units}
\end{equation}
We note that the operators in Eqs. \ref{ch2_eq:poissongk_units} and \ref{ch2_eq:Ampèregk_units} are written in Fourier space, e.g. $\nabla = i\bm k$, which simplifies their expression. 
Similarly, the gyro-averaging operator is expressed by a Bessel function of the first kind, $\Upsbar=J_0\Upsilon$.
% The Bessel function $J_0(a_\perp)$ expresses the transformation from gyrocentre to particle coordinates in Fourier space, i.e.
% \begin{equation}
    % \int \bar f_k e^{i\bm k \cdot \bm R} d\bm k= \int J_0(a_\perp) \bar f_k e^{i\bm k \cdot \bm x},
% \end{equation}
% for a generic Fourier mode of a gyro-averaged quantity $\bar f_k$, where we used the first order approximation $\int e^{\bm k \cdot \bm \rho_a} d\alpha = J_0(\kperp \vperp/\Omega_a)$ and the local approximation.
The collision operator used in this work is the GK Dougherty operator \citep{Dougherty1964, Frei2021DevelopmentApproach}.
This operator is identified as appropriate in conditions far from marginal stability and at sufficiently large collisionality \citep{Hoffmann2023GyrokineticOperators}.

 % \subsection*{Moment-based approach}

In this work, we use the \gyacomo code \citep{gyacomo}, to solve the GK equations, Eqs. \ref{eq:3d_gyboeq_nondim}-\ref{ch2_eq:Ampèregk_units}.
\gyacomo is based on the GM approach, that is the projection of the distribution function on a Hermite-Laguerre basis in the velocity space \citep{Mandell2018, Jorge2019b, Frei2020}. 
The GMs are defined as
\begin{equation}
    N_a^{pj}(k_x,k_y,z,t) = \iint \mathrm d x \mathrm d y  \iint\mathrm d\wperpa \mathrm d\spara  g_a H_p(\spara) L_j(\wperpa) e^{-ik_x x - ik_y y},
    \label{eq:GMs}
\end{equation}
where $k_x$ is the radial wavenumber, $k_y$ the binormal wavenumber, $H_p$ the normalized physicist's Hermite polynomial of order $p$, and $L_j$ the Laguerre polynomial of order $j$.
Here, the local limit allows for periodic boundary conditions in the perpendicular direction.
\gyacomo evolves Hermite-Laguerre-Fourier modes of the distribution function, which can be obtained by using the orthonormal properties of the Hermite-Laguerre basis as
\begin{equation}
    g_a(k_x,k_y,z,\spara,\wperpa,t) = \sum_{p=0}^\infty\sum_{j=0}^\infty N_a^{pj}(k_x,k_y,z,t) H_p(\spara)L_j(\wperpa).
    \label{eq:ga_projection}
\end{equation}
Projecting the GK system, Eqs. \ref{eq:3d_gyboeq_nondim}-\ref{ch2_eq:Ampèregk_units}, on the Hermite-Laguerre basis, as in Eq. \ref{eq:GMs}, we obtain the EM nonlinear GM hierarchy in a flux tube configuration \citep{Frei2023Moment-basedModel, Hoffmann2023GyrokineticShift, Mandell2023GX:Design},
\begin{equation}
    \ddt N_a^{pj} + \mathcal S_a^{pj} + \mathcal M_{\perp a}^{pj} + \mathcal M_{\parallel a}^{pj} + \mathcal D_{Na}^{pj} + \mathcal D_{Ta}^{pj} = \sum_b \nu_{ab}\mathcal C_{ab}^{pj},
    \label{eq:moment_hierarchy}
\end{equation}
where $\mathcal S_a^{pj}$ represents the projection of the nonlinear $\ExB$ advection, $\mathcal M_{\perp a}^{pj}$ the magnetic perpendicular gradient and curvature effect, $\mathcal M_{\parallel a}^{pj}$ the mirror force and Landau damping, $\mathcal D_{Na}^{pj}$ the density diamagnetic term, $\mathcal D_{Ta}^{pj}$ the temperature diamagnetic term, and $\mathcal C_{ab}^{pj}$ the projection of the collision operator.
The terms in Eq. \ref{eq:moment_hierarchy} are presented in \ref{app:emgmhierarchy} and more details on the GM model can be found in \cite{Frei2023Moment-basedModel} and \cite{Hoffmann2023GyrokineticShift}.
The GM hierarchy, Eq. \ref{eq:moment_hierarchy}, is closed by using a truncation closure, i.e. we impose $N_a^{pj}=0$ for $p>P$ and $j>J$, therefore $P$ and $J$ are the maximal degree of the considered Hermite and Laguerre polynomial basis, respectively.
In the following, we denote a finite set of GMs by the doublet $(P,J)$.

In this work, we compare three different levels of fidelity, all at the ion-scale spatial resolution, i.e. we evolve Fourier modes up to $k_{y}\rho_s \sim 1$.
First, we consider a kinetic description for the ions and for the electrons, which we denote as the kinetic electron model (KEM).
This model solves the electromagnetic GK Boltzmann equation, Eq. \ref{eq:3d_gyboeq_nondim}, for both the ion and electrons species, with the quasi-neutrality equation, Eq. \ref{ch2_eq:poissongk_units}, and the Ampère equation, Eq. \ref{ch2_eq:Ampèregk_units}.
% We use a \PJ{4}{1} GM basis, which is proved to be sufficient far from marginal stability of the ITG \citep{Hoffmann2023GyrokineticShift}.
% The ion-scale spatial resolution prevents the destabilization of the ETG in these simulations.
% In fact, the peak growth rate of the ETG instability occurs at $k_y\rho_e\sim 0.5$ and turbulence presents an inverse cascade, where fluctuations with larger wavelengths dominate \citep{Rogers2005, Howard2016Multi-scaleTransport}.
Second, we consider the electrostatic adiabatic electron model (AEM), where the electron GK equation is not evolved and, by using the adiabatic electron approximation, the quasi-neutrality equation, Eq. \ref{ch2_eq:poissongk_units}, becomes
\begin{equation}
     \left( 1 + \frac{q_i^2}{\tau_i}\left[1-\sum_{n=0}^{\infty}\left(\kernel_{i}^n\right)^2\right]\right)\phi - \langle \phi \rangle_{yz} = q_i\sum_{n=0}^{\infty}\kernel_{i}^n N_i^{0n},
    \label{ch5_eq:poisson_moments_adiabe}
\end{equation}
where $\langle \phi \rangle_{yz}$ is the flux surface average of $\phi$, namely
\begin{equation}
    \langle \phi \rangle_{yz} = \frac{1}{\int\dz J_{xyz}}\int\dz  J_{xyz}\phi(k_x,k_y,z,t)\delta_{k_y0}.
\end{equation}
In this case, electromagnetic fluctuations are neglected, which sets $A_\parallel  = 0$.
% Similarly to the KEM, the AEM uses a \PJ{4}{1}.
Finally, we consider a reduced fluid model (RFM) for ITG-driven turbulence, composed of a set of 4 GMs ($N_i^{00}$, $N_i^{10}$, $N_i^{20}$, $N_i^{01}$) with an adiabatic and hot electron asymptotic closure.
This model is equivalent to the one in \cite{Ivanov2022DimitsTurbulence}, and is obtained by setting the ion-electron temperature ratio to $\tau=10^{-3}$ in the \gyacomo code, scaling the temperature background gradients using $\kappa_T=R_T/2\tau$, and the collision rate with $\chi=8\nu/3\tau$ \citep{Hoffmann2024thesis}.

\section{Experimental profiles and numerical setup}
\label{sec:exp_prof_and_num}

\begin{figure}
    \centering
    \includegraphics[width=0.61\linewidth]{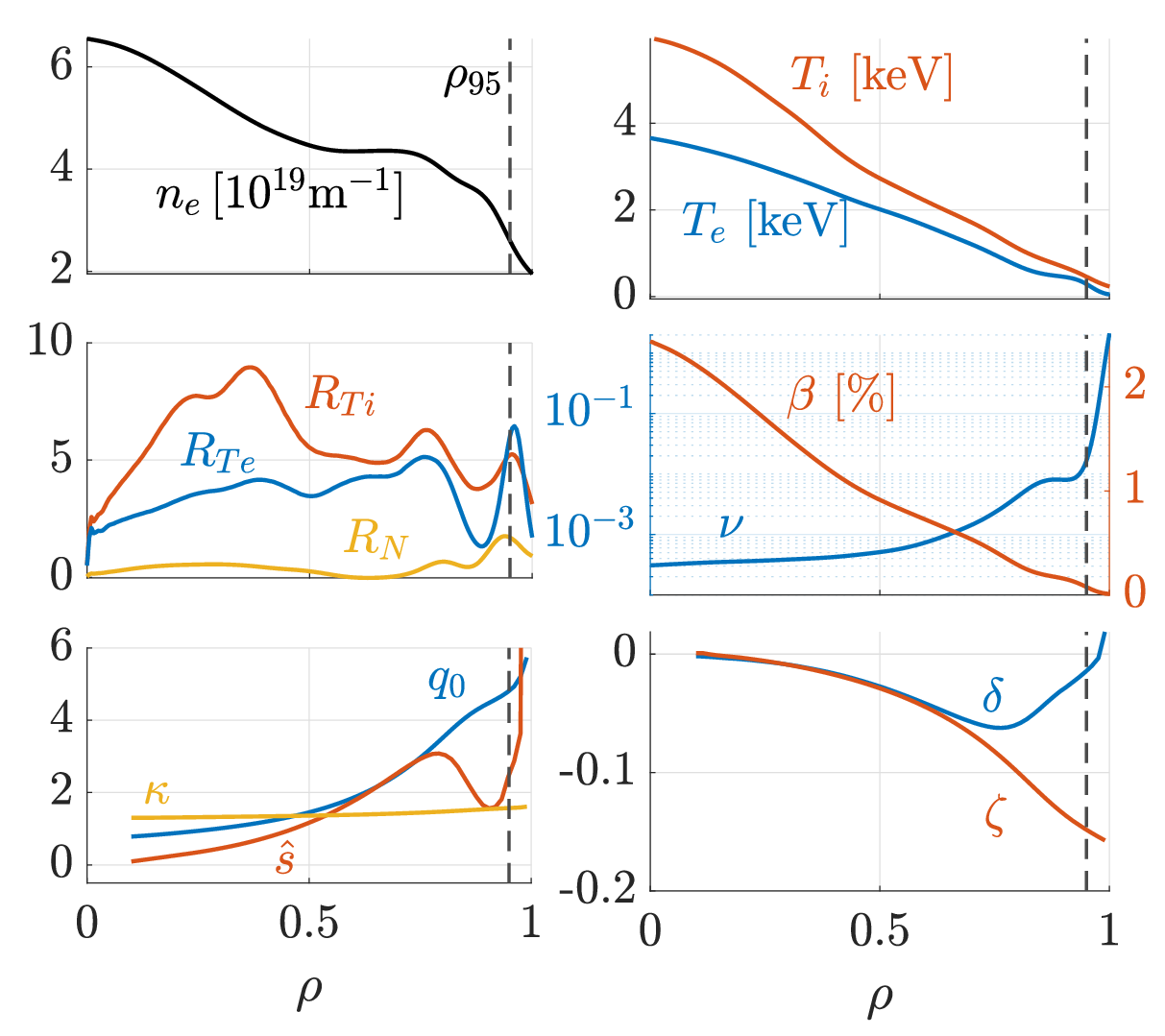}
    \includegraphics[width=0.38\linewidth]{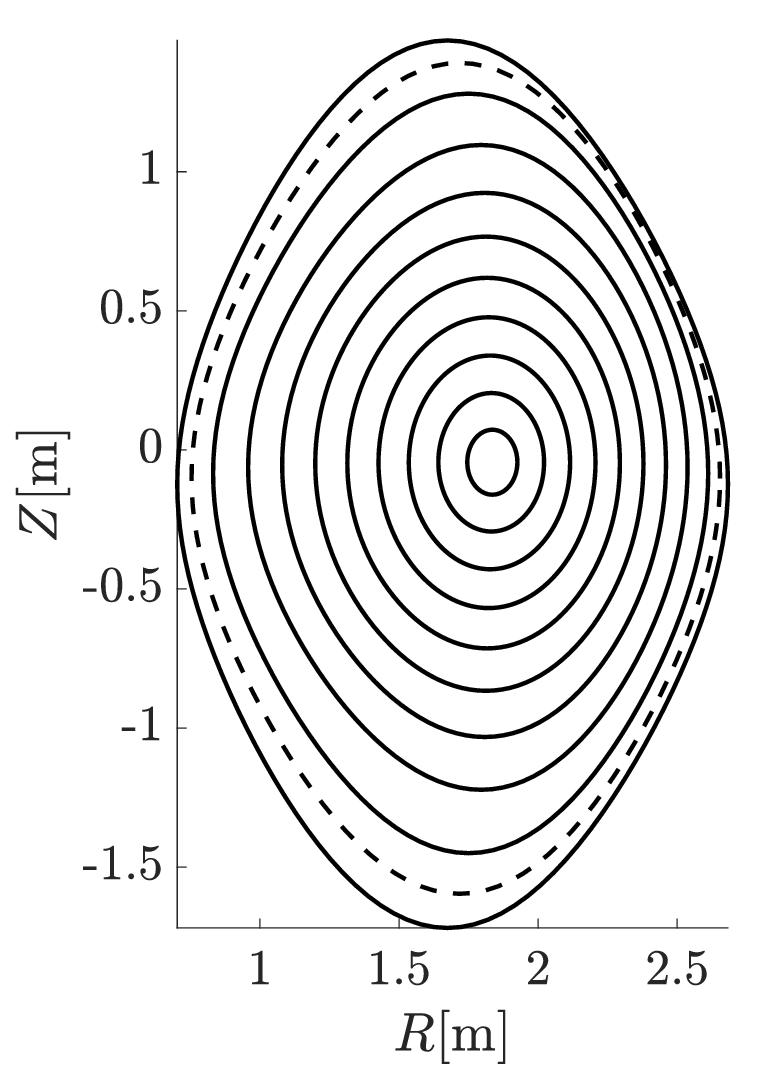}
    \caption{Profiles of the DIII-D discharge \#186473 at $t=3500$ms. The dashed black lines indicate the $\rho=0.95$ position, considered in the GK simulations. 
    Error bars are not displayed here but are usually of the order of 10\%.}
    \label{ch7_fig:DIII-D_shot}
\end{figure}

To investigate the influence of triangularity on edge turbulence, we consider the parameters of the DIII-D discharge \#186473 presented in \cite{Boyes2023MHDPlasmas}, where the MHD stability properties of this configuration are investigated.
This discharge was employed experimentally to assess the impact of shaping on a broad range of parameters, motivating its use for our investigation. 
The reconstructed profiles of the discharge are shown in Fig. \ref{ch7_fig:DIII-D_shot} and, for our simulations, we consider the parameters at $\rho=0.95$ (see Tab. \ref{ch7_tab:DIIID_params}).

The considered DIII-D discharge features an L-mode plasma. 
The edge profile gradients are close to the cyclone base case (CBC) \citep{Lin1999EffectsTransport,Dimits2000ComparisonsSimulations}, therefore far from the marginal stability conditions of the ITG and ETG modes \citep{Jenko2001CriticalModes,Hoffmann2023GyrokineticShift} and, as a consequence, from the Dimits shift.
However, the safety factor is significantly enhanced with respect to the CBC, primarily due to the larger local aspect ratio, $\rho/R_0$, and the smaller poloidal field.
The large local variation of the safety factor yields a strong magnetic shear, challenging the flux-tube representation numerically.
Indeed, the minimum radial wavenumber to consider (i.e., the largest wavelength) in a flux-tube representation is proportional to the shear value \citep{Ball2020}.
%Thus, to accommodate for a perpendicular plane of dimension $L_x\sim L_y\sim 80\rho_s$, the $N_{exc}$ must be increased, yielding a large number of disconnected $k_y$ modes\footnote{The $j$-th $k_y$-mode is defined between $\pm (m+1)\pi$ in the ballooning space representation if $j+mjN_{exc}\leq N_x$ with $m$ a positive integer. 
%Here, $m$ represents the number of connections to other existing modes through the parallel shear boundary conditions and we consider it disconnected if $m=0$.}.
Consequently, the edge magnetic parameters require an increased number of radial wavenumbers in comparison to the CBC parameters.
Finally, we note that the plasma $\beta$ value is low, thus justifying the assumption $\delta B_\parallel=0$, and the collision frequency is ten times larger than in the CBC.

In our simulations, we consider an artificially increased electron-to-ion mass ratio, $m_e/m_i = 1/1000$, effectively decoupling the ETG and ITG instability scales by a factor of ten against twenty for a realistic deuterium plasma. 
This allows simulations to run with a time step $\Delta t=5\times 10^{-3}$.
%As in \cite{Belli2023SpectralPedestal}, we neglect the density gradient to simplify the analysis, noting that its experimental value is within the error margin.
%Following the same argument, we also neglect the squareness.
Although the plasma beta value is low, we evolve the Ampère equation, Eq. \ref{ch2_eq:Ampèregk_units}, for the parallel component of the magnetic vector, confirming the absence of dominating instabilities with an EM character.

%For the multi-scale simulations, we use a \PJ{2}{1} GM basis, which is sufficient to simulate highly turbulent systems. 
%the flux-tube domain is composed of $N_z=24$ perpendicular planes of size $L_x=L_y=90$ with resolution of $N_x=768$ and $N_y=384$. 
%This yields a grid spacing of 3$\rho_e$, which is sufficient to resolve the ETG peak growth rate and the ETG-driven turbulence.
%In fact, the peak growth rate of the ETG instability occurs at $k_y\rho_e\sim 0.5$ and turbulence presents an inverse cascade, where fluctuations with larger wavelengths dominate \citep{Rogers2005, Howard2016Multi-scaleTransport}. 
%To suppress the unresolved portion of the unstable ETG spectrum, we set the spatial hyperdiffusion parameter to $\mu=1.0$.

% The RFM does not retain the effect of the density background gradient.\\
All our simulation setup are evolving ion-scale turbulence, i.e. evolving modes up to $k_y\rho_s \sim 1$. 
The nominal spatial resolution parameters are $N_x=256$, $N_y=64$, and $N_z=24$.
The KEM and AEM use a flux tube of dimension $L_x=100$ and $L_y=120$ perpendicular to the magnetic field, while the RFM considers $L_x=150$ and $L_y=300$ due to the larger size of the turbulent eddies observed in the RFM simulations.
These numerical parameters guarantee the numerical convergence of the simulations we present, as they are chosen to be $50\%$ larger than the minimal resolution required to obtain an accurate nonlinear transport level in the considered scenario.

To study the impact of triangularity, the $\delta=0$ parameter of the DIII-D discharge \#186473, is varied to also consider $\delta =0.2$ and $-0.2$ for PT and NT scenarios, respectively.
% We compare the resulting stationary transport level also among our simulations with different levels of fidelity.
This range of triangularity values considered is also explored in the experimental scenario of \cite{Boyes2023MHDPlasmas}, while maintaining the plasma profiles unchanged.
In contrast, optimized H-mode discharges often operate within a limited triangularity window and would not be suitable for the parameter scan discussed here.

\begin{table}
    \centering
    \begin{tabular}{l r l| l r l}
    Safety factor     & $q_0$ & $= 4.8$  
    &%\\
    Density gradient    & $R_N$ & $= 1.7$ 
    \\
    Magn. shear & $\hat s$ & $= 2.5$
    &%\\
    Electr. temp. gradient & $R_{T_e}$ & $= 6.0$
    \\
    Inverse aspect ratio & $\epsilon$ & $= 0.3$ 
    &%\\
    Ion temp. gradient & $R_{T_i}$ & $= 5.2$ 
    \\
    Elongation & $\kappa$ & $= 1.6$ 
    &%\\
    Temperature ratio & $T_i/T_e$ & $= 1.6$
    \\
    Elongation shear & $\hat  s_\kappa$ & $= 0.5$ 
    &%\\
    % Electron effective collision & $\nu_e^*$ & $= 1.4$
    Mass ratio & $m_i/m_e$ & $= 10^3\, (3.7\times10^3)$
    \\
    Triangularity & $\delta$ & $= 0.0 $
    &%\\
    Collision parameter & $\nu$ & $= 1.7\times 10^{-2}$
    \\
    Triangularity shear & $s_\delta$ & $=0 (0.1)$
    &%\\
    Magn. pressure ratio & $\beta$ & $= 7.6\times 10^{-4}$ 
    \\
    Squareness & $\zeta$ & $=0 \,(-0.15) $
    & 
    \end{tabular}
    \caption{Normalized parameters considered for the GM simulations, obtained from the equilibrium of the \#186473 NT DIII-D discharge at $\rho=0.95$. When a parameter is adapted, we show its real value in parenthesis.}
    \label{ch7_tab:DIIID_params}
\end{table}

\section{Gyrokinetic edge simulations of the DIII-D discharge \#186473}
\label{sec:nominal_parameters_simulations}
We present GK simulations based on the \gyacomo code that consider the nominal parameters of the DIII-D discharge \#186473, as listed in Tab. \ref{ch7_tab:DIIID_params} and, in particular, $\delta =0$.
We aim to identify the instabilities driving the turbulent dynamics and evaluate the number of GMs required for convergence in linear and nonlinear simulations.

\begin{figure}
    \centering \includegraphics[width=0.5\linewidth]{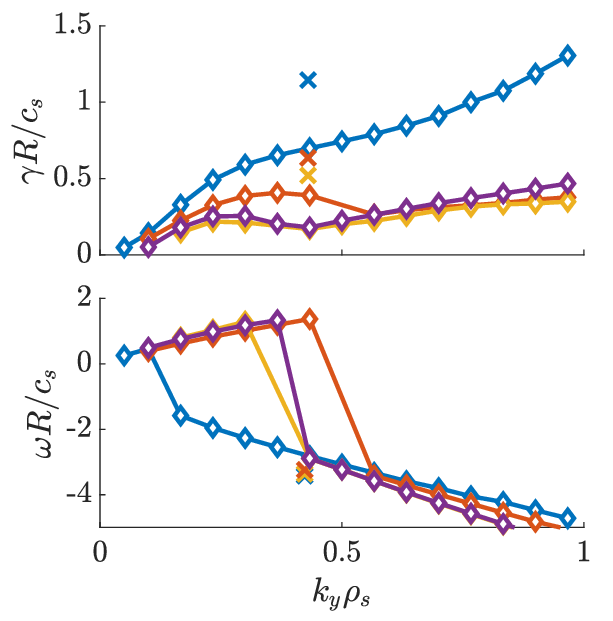}
    \caption{Linear growth rates (top) and frequencies (bottom) for the nominal parameters in Tab. \ref{ch7_tab:DIIID_params} with different GM sets: \PJ{2}{1} (blue line), \PJ{4}{2} (red line), \PJ{8}{4} (yellow line), and \PJ{16}{8} (purple line).}
    \label{fig:linear_results}
\end{figure}

% \subsection*{Linear results}
We first focus on the properties of the linear instabilities.
Figure \ref{fig:linear_results} presents scans of the growth rates $\gamma$ and frequencies $\omega$ as a function of $k_y$, having set $N_x=8$ and $N_z=24$.
%obtained with the amplitude ratio method \citep{Frei2023Moment-basedModel}.
At long wavelength ($k_y\rho_s \leq 0.4$), the ITG mode is the fastest growing instability.
This is identified by the positive frequency, propagating in the ion direction.
At shorter wavelengths ($k_y\geq 0.4$), we find that the TEM instability dominates, in good agreement with previous investigations \citep{Happel2023OverviewTokamak,Merlo2023InterplayPlasmas}.
We note that neglecting magnetic mirror and trapping terms (the terms proportional to $\ddz\ln B$ in Eq. \ref{ch2_eq:Mparapj}) leads to a stabilization of the TEM instability, without affecting the ITG mode, confirming the nature of these modes.

We also study the convergence of our linear results with respect to the number of GMs.
The unstable region of the TEM and its growth rate are overestimated when the \PJ{2}{1} GM basis is used.
On the other hand, the $(4,2)$ GM basis predicts qualitatively good linear results with a transition between ITG and TEM occurring at $k_y\rho_s \sim 0.5$, which is larger than the converged results.
This is mostly due to an overestimate of the ITG growth rates for this particular mode number.
In fact, when the ion temperature gradient is set to zero, we retrieve a pure TEM instability for the $(4,2)$ and $(8,4)$ sets, with similar growth rates and frequencies.
Finally, only small differences are observed between the $(8,4)$ and $(16,8)$ GM results. 
These differences increase with $k_y$ as previously observed \citep{Hoffmann2023GyrokineticOperators,Frei2023Moment-basedModel,Hoffmann2023GyrokineticShift}.
% Consequently, we observe that the convergence is set by the resolution of the ITG instability.
% This indicates that the core mechanisms of the TEM instability rely here on the destabilizing role of the density gradient, which introduces energy in the density GM, namely $N_a^{00}$.
% The coupling with neighbouring GMs, parallel velocity GM ($N_a^{10}$) and temperature GMs ($N_a^{20}$ and $N_a^{01}$) is then possible even with the $(2,1)$ basis.
% On the other hand, the ITG mode is destabilized by the temperature gradient which injects energy into the temperature GMs.
% In this case, the $(2,1)$ basis cannot resolve the direct coupling between the temperature GMs and higher-order GMs, altering strongly the result.

%This ETG unstable is validated by \cite{Jenko2001CriticalModes}.

\begin{figure}
    \centering
    \includegraphics[width=1.0\linewidth]{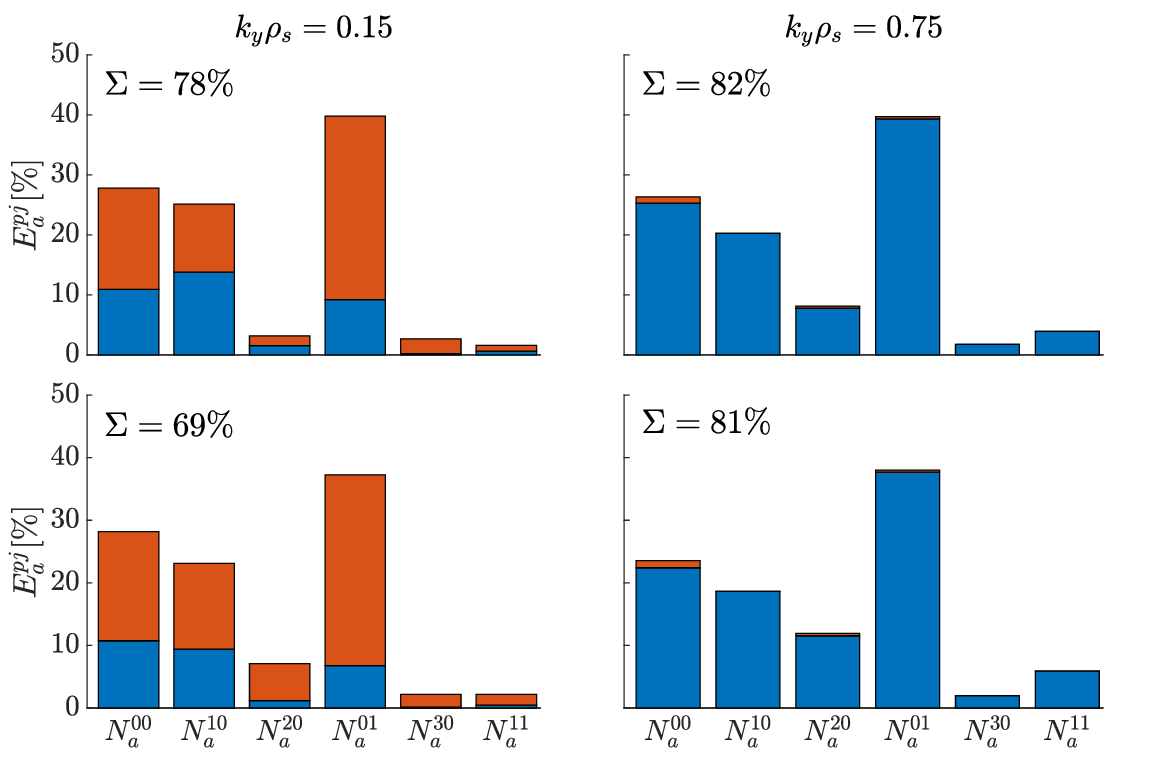}
    \caption{Amplitude of the electron (blue) and ion (red) low-order GMs, $E_a^{pj}$, with the \PJ{4}{2} basis (top) and \PJ{16}{8} basis (bottom) normalized to $E_{tot}$ for $k_y\rho_s=0.15$ (left) and $k_y\rho_s=0.75$ (right).
    The ratio between the sum of the moment amplitudes presented here, versus the total amplitude is represented by $\Sigma=\sum_a(N_a^{00}+N_a^{10}+...+N_a^{11})/E_{tot}$.}
    \label{fig:linear_energy_fluid_distribution}
\end{figure}

For a more detailed convergence analysis, we compute the relative amplitude of the GMs, more precisely the sum of the squared modulus of the Fourier modes for a given GM, $E_a^{pj} = \sum_{k_x} |N_a^{pj}|^2$,
normalized to the sum of all GMs, $E_{tot}=\sum_{a,p,j} E_a^{pj}$.
This is shown in Fig. \ref{fig:linear_energy_fluid_distribution}, where the GM amplitudes of the fastest linear instability are compared between the $(4,2)$ and the $(16,8)$ sets, considering two different wavenumbers.
The agreement with the $(16,8)$ set confirms the capability of the $(4,2)$ set to reproduce ITG and TEM for the considered parameters.
Discrepancies appear only in the wavenumbers where both the ITG and TEM have similar growth rates ($k_y\rho_s\sim 0.5$, not shown), confirming that the transition between these two instabilities is not accurately predicted by the $(4,2)$ set.
We also evaluate the importance of the higher-order GMs by comparing the sum of the amplitude of the six GMs presented in Fig. \ref{fig:linear_energy_fluid_distribution} with the sum of the amplitudes of all GMs.
For the ITG case ($k_y\rho_s=0.15$), the ion GMs have a larger amplitude than the electron ones, and the $(16,8)$ set shows a larger importance of higher-order GMs, $30\%$, against $20\%$ for the $(4,2)$ set.
In contrast, the electron GMs have a larger amplitude in the TEM-dominated regime ($k_y\rho_s=0.75$) and the considered lower-order GMs represent $80\%$ of the total amplitude for both sets.
This supports the observation that the TEM growth rate can be efficiently evaluated with a reduced set of GMs, even if the fine features of the passing-trapping boundary in the velocity space require $\sim 100$ GMs for a proper description \citep{Frei2023Moment-basedModel}.
Indeed, the TEM instability is driven by the resonance between a drift wave and the bouncing frequency of the trapped electron population.
The accurate resolution of the passing-trapping boundary concerns a minority of the resonant particles, thus having a minor effect on the growth rate.

\begin{figure}
    \centering
    \includegraphics[width=\linewidth]{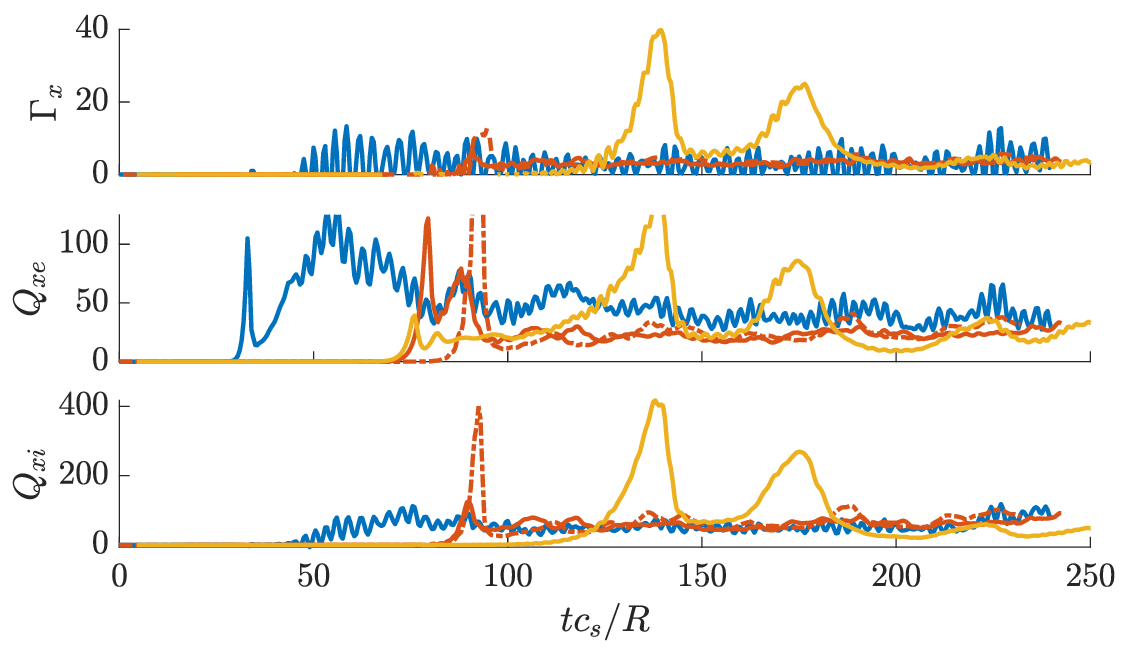}
    \caption{Time evolution of the electron (top) and ion (bottom) heat fluxes for the simulations of discharge \#186473, with nominal parameters in Tab. \ref{ch7_tab:DIIID_params}. 
    Different GM sets are considered: \PJ{2}{1} (blue line), \PJ{4}{2} (red line), \PJ{8}{4} (yellow line). The dashed lines represent simulations with a spatial resolution of $192\times 48 \times 24$.}
    \label{fig:nonlinear_convergence_0T}
\end{figure}

% \subsection*{Nonlinear results}
We now turn to the nonlinear simulation of the DIII-D discharge \#186473. 
% of the parameters presented in Tab. \ref{ch7_tab:DIIID_params} and evaluate the convergence of the heat transport with respect to the number of GMs evolved.
Figure \ref{fig:nonlinear_convergence_0T} presents the electron and ion radial heat fluxes as a function of time, comparing simulations with the \PJ{2}{1}, $(4,2)$, $(8,4)$ basis, as well as a lower spatial resolution run. 
Since the simulations are initialized with a small amplitude noise in the electrostatic potential, the system first shows an exponential growth of the unstable modes.
Then, these saturate because of the growth of a secondary Kelvin-Helmholtz instability (KHI) \citep{Rogers2005}.
The time-averaged particle and heat fluxes (obtained excluding the initial transient) are reported in Tab. \ref{tab:nominal_parameters_averages}.
The $(2,1)$ simulation overestimates the electron heat flux, a phenomenon that can be related to the enhanced TEM growth rate, as observed in the linear case (see Fig. \ref{fig:linear_results}).
On the other hand, the saturated level of the ion heat flux observed in the $(2,1)$ simulation is close to the one observed in the other simulations, showing that ITG-driven turbulence is resolved even within a very small number of moments.
We remark that large oscillations are observed in the heat flux of both species in the $(2,1)$ simulations.
These oscillations, not observed in the simulations carried out with a larger number of moments, are due to the inaccurate resolution of the Landau damping.
This results in spurious time-oscillating zonal structures that suppress turbulence \citep{Hoffmann2024thesis}.
We also note the close agreement between the two simulations carried out with different spatial resolutions.

We note that the transient period of the $(8,4)$ simulation differs from the other simulations.
This is due to the discrepancies in the ITG growth rates observed for this set of GMs, as shown in Fig. \ref{fig:linear_results}.
However, the saturated transport value of the $(8,4)$ simulation is very close to the $(4,2)$ one.
In fact, similarly to the CBC studied in \cite{Hoffmann2023GyrokineticShift}, the number of moments necessary for the convergence of the nonlinear saturated transport is reached with a smaller number of GMs than the convergence of the linear growth rates.

\begin{table}
    \centering
    \begin{tabular}{cccc}
        $(P,J)$ & $\langle \Gamma_x \rangle_t$ &$\langle Q_{xi} \rangle_t$ &$\langle Q_{xe} \rangle_t$ \\\hline
       $(2,1)$ & $2.9 \pm 0.8$ & $54.8 \pm 12.9$ & $42.3 \pm 6.4$ \\
       $(4,2)^*$ & $3.6 \pm 0.8$ & $64.2 \pm 15.9$ & $23.0 \pm 5.5$ \\
       $(4,2)$ & $3.3 \pm 0.3$ & $60.5 \pm 7.3$ & $23.0 \pm 2.4$ \\
       $(8,4)$ & $4.0 \pm 0.5$ & $63.5 \pm 10.1$ & $21.4 \pm 3.1$ 
    \end{tabular}
    \caption{Time-averaged radial particle flux, $langle \Gamma_x \rangle_t$, and the ion and electron radial heat fluxes, $\langle Q_{xi} \rangle_t$ and $\langle Q_{xe} \rangle_t$, respectively. 
    The $(4,2)^*$ set denotes the simulation with a spatial resolution of $192\times 48 \times 24$.}
    \label{tab:nominal_parameters_averages}
\end{table}

\begin{figure}
    \centering
    \includegraphics[width=\linewidth]{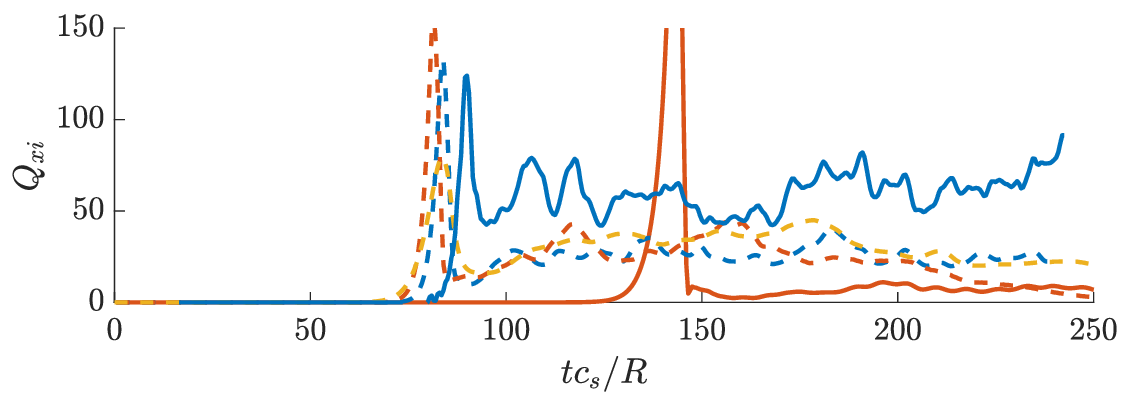}
    \caption{Time traces of the ion heat flux obtained with the KEM (blue line), AEM (red line), and RFM (yellow line), considering the nominal parameters of discharge \#186473 (solid line) and neglecting the density gradient, $R_N=0$ (dashed line).}
    \label{fig:HF_nominal_vs_nogradN_KE_vs_AE}
\end{figure}
% \section{Comparison of the different models}
% \label{sec:model_comparison}
We finally focus on the \PJ{4}{2} GM set and compare the transport predictions obtained by this simulation (KEM), with the prediction of AEM and RFM simulations.
Nonlinear simulations based on these models are performed and the resulting ion heat fluxes are presented in Fig. \ref{fig:HF_nominal_vs_nogradN_KE_vs_AE}.
The KEM predicts a considerably higher transport level than the AEM simulation where TEM are absent, since the adiabatic electron response cannot capture the effects of trapped particles.
Indeed, when the density gradient is zeroed out, stabilizing the TEM in the KEM simulation \citep{Adam1976DestabilizationResonances}, we observe a similar transport level in the KEM and AEM simulations.
At the same time, the turbulent transport level increases because of the stabilizing role of the density gradient on the ITG instability \citep{Mikhailovskii1974TheoryInstabilities} and a good agreement also with the RFM simulation is observed, confirming the ITG nature of turbulence in this setup.

% Additionally, the timing of the KHI between the KEM and AEM simulations is very similar, showing that both models are dominated by the same ITG instability.

\begin{figure}
    \centering    
    \includegraphics[width=0.9\linewidth]{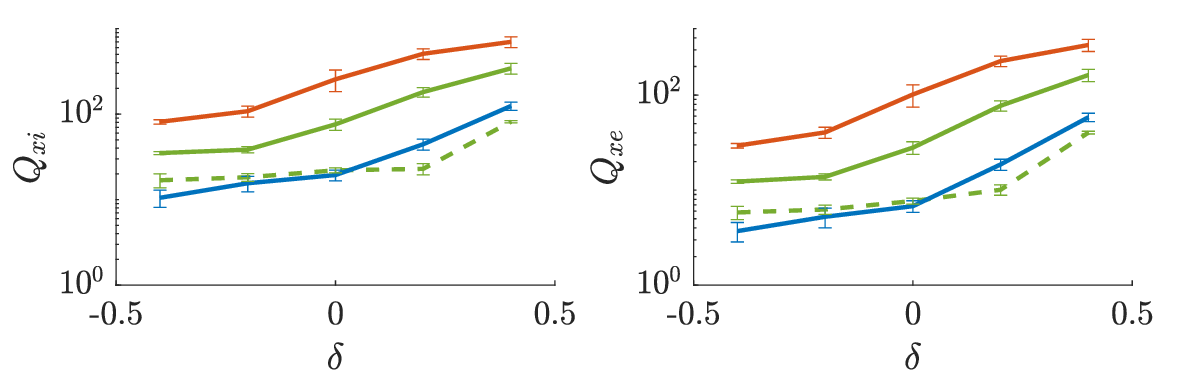}
    \caption{Time-averaged radial heat fluxes for ion (left) and electrons (right) obtained with KEM simulations using the nominal density and temperature gradient values (green line), a $25\%$ increase of these values (red line) and a $-25\%$ decrease (blue line). The dashed line is obtained by using the nominal parameters, but zeroing out the density gradient.}
    \label{fig:heat_flux_KEM}
\end{figure}
\section{Impact of triangularity on turbulent transport}
\label{sec:triangularity_scan}

We now investigate the impact of triangularity on the saturated turbulent heat flux levels by scanning $\delta$ while assuming a triangularity shear of $s_\delta = -\delta/2$.
Figure \ref{fig:heat_flux_KEM} presents the saturated turbulent heat flux level obtained with KEM simulations with different values of triangularity. 
When considering the nominal background gradient values, we observe a monotonic increase in both ion and electron heat flux with increasing triangularity.
However, the sensitivity of the transport to triangularity is reduced for $\delta \leq -0.3$.
We also examine the sensitivity of our results to a $\pm 25\%$ variation of the background density and temperature gradients.
In both cases, the transport shows similar trends.
On the other hand, when the density gradient is zeroed out, transport shows a weaker sensitivity to triangularity for $\delta < 0.2$.
This behavior is a consequence of the sensitivity of the TEM instability to the values of the background gradients and the triangularity, while a smaller sensitivity is shown by the ITG instability. 
In fact, when the sensitivity to triangularity is high, turbulence is driven by TEMs, which are highly sensitive to the magnetic geometry as their destabilization relies on trapping effects.
In the TEM turbulent regime, a reduction of triangularity yields a reduction of the fluctuation amplitude, thus improving the confinement.
On the other hand, once the TEMs are stabilized, as a consequence of either the triangularity or the gradients values \citep{Merlo2023InterplayPlasmas}, the transport level is set by the ITG instability, which is less sensitive to the magnetic geometry.
Hence, the confinement improvement related to the reduction of the triangularity observed in current tokamaks, which are often TEM-dominated, may be less evident in future larger machines, such as ITER, where the transport is expected to be driven by the ITG.

\begin{figure}
    \centering    
    \includegraphics[width=0.9\linewidth]{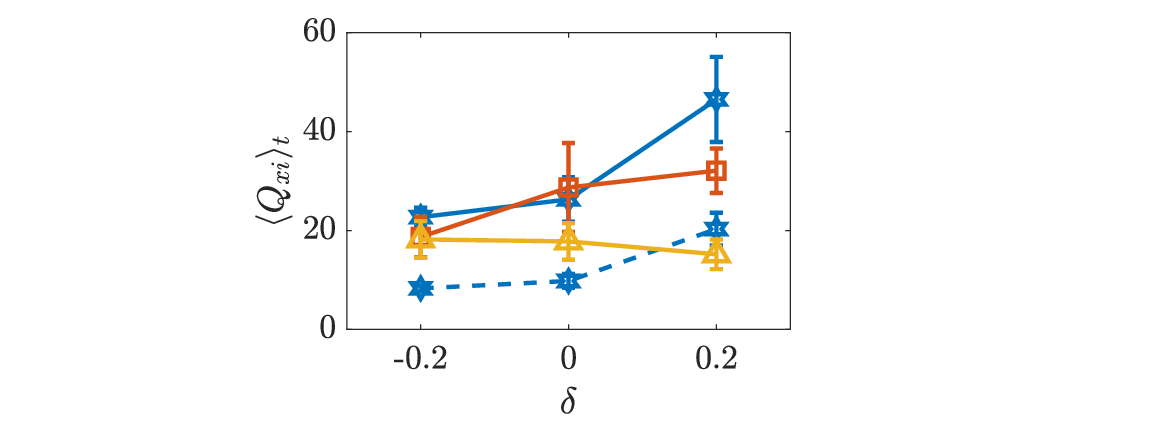}
    \caption{Time-averaged radial ion heat fluxes obtained with the KEM (solid blue line), AEM (solid red line), and RFM (solid yellow line) simulations.
    The electron heat fluxes for the kinetic electron simulations are also presented (dashed blue line).
    The error bars represent the standard deviation.
    Here, the density gradient is neglected, the other simulation parameters are taken from Tab. \ref{tab:nominal_parameters_averages}.}
    \label{ch7_fig:heat_flux_iscale_PT_0T_NT}
\end{figure}

To study future machine scenarios and compare our different models, we now focus on ITG-dominated turbulence, by zeroing out the density gradient and considering $|\delta| \leq 0.2$.
Figure \ref{ch7_fig:heat_flux_iscale_PT_0T_NT} presents the time-averaged ion heat flux obtained from nonlinear KEM, AEM, and RFM simulations, for three values of triangularity, namely $\delta=-0.2$ (NT), $\delta=0.0$ (0T), and $\delta=0.2$ (PT).
We observe, first, that the 0T and NT cases present roughly similar transport levels.
The AEM and KEM predict an increase of the transport level for the PT scenario in comparison to 0T and NT, with a stronger increase in the KEM simulation, which is due to the role of TEMs.
On the other hand, although the RFM shows reasonable agreement in heat flux levels between the NT and 0T cases, it fails to predict the transport increase in the PT configuration. 

\begin{figure}
    \centering
    \includegraphics[width=\linewidth]{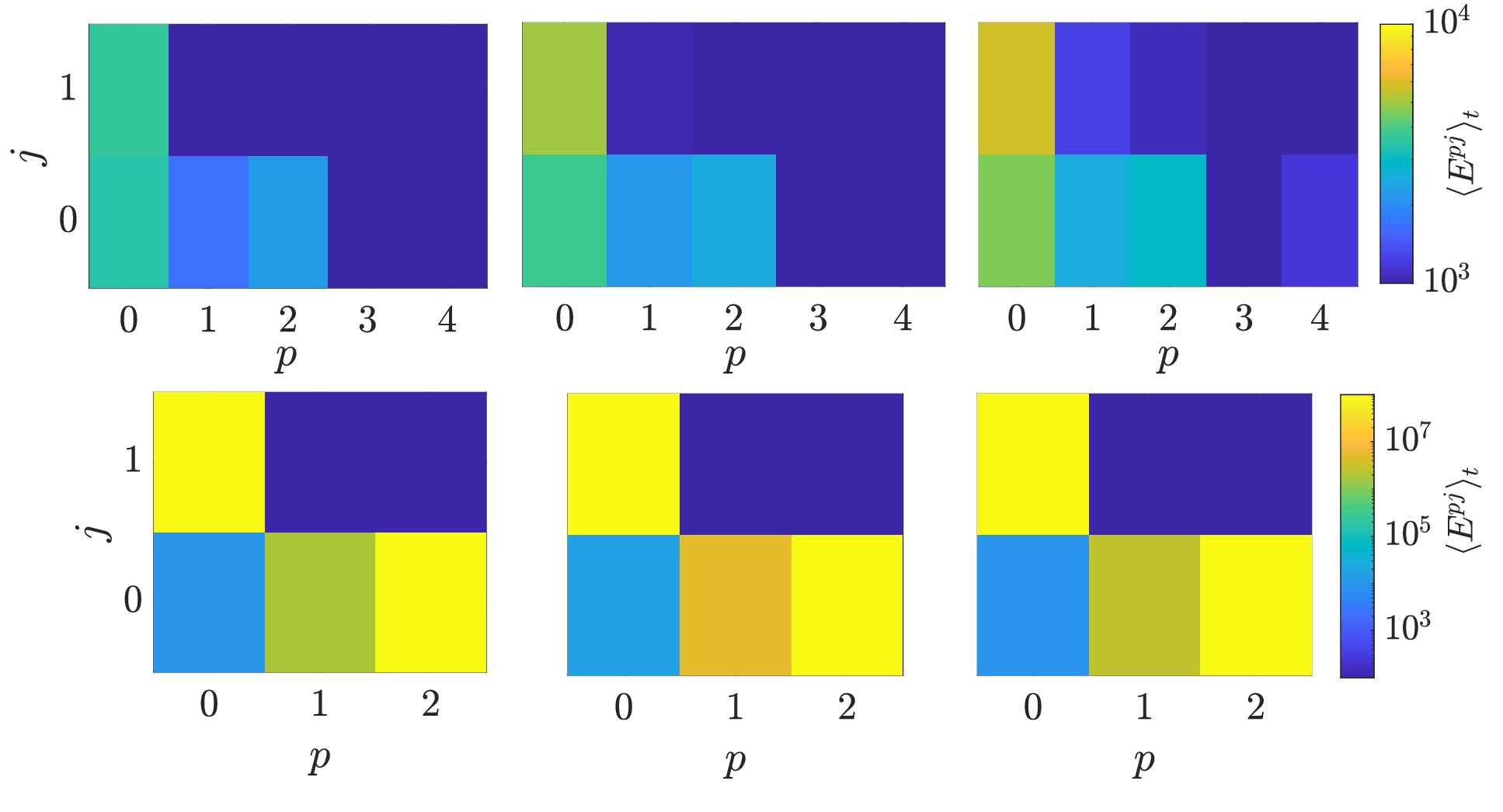}
    \caption{Time-averaged amplitude of the GMs during the nonlinear saturated phase for the AEM (top) and the RFM (bottom) considering NT (left), 0T (middle), and PT (right).}
    \label{ch7_fig:adiabe_HEL_Epj_comparison}
\end{figure}

The increase of the heat flux in the PT case, shown by the AEM simulation, in contrast to the RFM simulation, suggests that triangularity affects the ITG mode through a kinetic mechanism that is not accounted for by the RFM.
This is confirmed by Fig. \ref{ch7_fig:adiabe_HEL_Epj_comparison}, where the amplitude of the GMs in the AEM and RFM simulations are compared for the considered three triangularity values.
The amplitude of the temperature-related GMs ($N_i^{20}$ and $N_i^{01}$) is several orders of magnitude larger in the RFM simulations than in the AEM ones, and a similar consideration can be done for the parallel velocity GM, $N_i^{10}$. 
We observe a small variation of the GM amplitude across triangularity, except for the parallel velocity, which appears to increase slightly for $\delta \neq 0$.
In contrast, the AEM simulations are characterized by the perpendicular temperature, $N_i^{01}$, and density, $N_i^{00}$, being the largest GMs. 
In addition, and even more interestingly, higher-order GMs ($p>2$), which are absent in the RFM, exhibit a noticeable increase in energy with increasing triangularity in the AEM simulations.
In fact, the RFM only retains first-order curvature terms (see Eq. \ref{ch2_eq:Mperpapj}) which are responsible for the excitation of the $N_i^{40}$ GM in the PT AEM simulation.
These findings highlight the limitations of reduced fluid models in capturing the interplay between kinetic effects and triangularity in ITG turbulence. 
%A comprehensive kinetic model is necessary to fully capture the impact of triangularity on heat transport in tokamak plasmas.

\begin{figure}
    \centering
    \includegraphics[width=\linewidth]{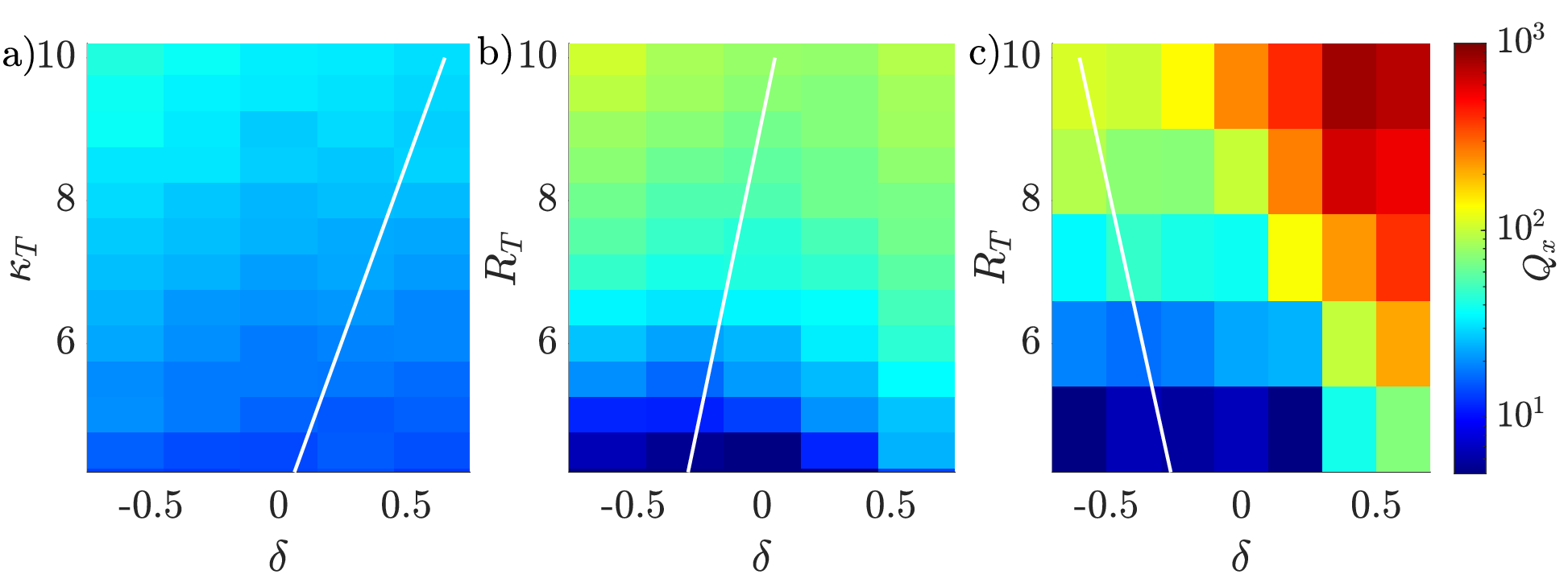}
    \caption{Time-averaged heat flux levels for the RFM (a), AEM (b), and KEM (c) simulations for different triangularity and temperature gradient values. The solid lines show the minimum of transport for a given temperature gradient, obtained through a second-order polynomial fit.
    The density gradient is neglected in these simulations, the other simulation parameters are given in Tab. \ref{tab:nominal_parameters_averages}.}
    \label{ch7_fig:triangularity_RT_scan_AES_HES}
\end{figure}

% \subsection*{Scan in temperature gradient}
The role of kinetic effects depends, however, on the temperature gradient.
This is illustrated in Fig. \ref{ch7_fig:triangularity_RT_scan_AES_HES}, where the saturated value of the heat flux is presented for different values of temperature gradients and triangularities for the RFM, AEM, and KEM simulations.
The value of $\delta$ that minimizes the heat flux for each gradient value is also highlighted.
We observe that the minimum heat flux value shifts from NT to PT as $\kappa_T$ increases in the case of the RFM simulations.
A similar dependence is observed in the AEM simulations.
% We note that the stiffness of the saturated heat flux with respect to temperature gradient is not identical, and the agreement between the heat fluxes of both models at $R_T=5$ appears like a coincidence.
Indeed, both models predict that NT does not suppress transport beyond a certain gradient value, an observation in agreement with \cite{Balestri2024PhysicalPlasmas}.
In addition, the AEM simulations recover the trend observed in previous GENE simulations \citep{Merlo2023OnTokamaks}, where both PT and NT increase the transport for large temperature gradients. 
This feature is not present in the RFM simulations, which predict a minimum of transport at high PT, highlighting a potential limitation of this reduced model.
On the other hand, a similar scan of KEM simulations reveals the importance of the TEM instability on the transport level. 
According to the KEM results, the NT configurations improve the confinement also when considering a strong temperature gradient ($R_T=10$), since the TEM is stabilized in these conditions.
However, when the TEM is stable, the KEM results agree closely with the AEM simulations, and a strong value of NT does not suppress the transport.
%\begin{figure}
%    \centering
%    \includegraphics[width=\linewidth]{figures/triangularity_RT_scan_AES_HES.eps}
%    \caption{Saturated heat flux levels for the AEM simulations (left) and the RFM simulations (right) for different triangularity and temperature gradient levels. The solid lines show the minimum of transport for a given temperature gradient, obtained through a polynomial fit.}
%    \label{ch7_fig:triangularity_RT_scan_AES_HES}
%\end{figure}
% \begin{figure}
%     \centering
%     \includegraphics[width=\linewidth]{figures/ion_scale_Epj_spectrum.eps}
%     \caption{Evolution of the GM amplitude, $E_{pj}=\sum_{k_x,k_y,z} |N^{pj}(k_x,k_y,z)|$, of ions (left) and electrons (right) during an ion-scale turbulence simulation of the DIII-D discharge \#186473 at $\rho=0.95$.}
%     \label{fig:enter-label}
% \end{figure}

%\begin{table}
%    \centering
%    \begin{tabular}{r|cccc}
%        $\delta$ & Multi-scale & Ion-scale & Adiab. electr. & Hot electr. \\ \hline
%          $ 0.2$ &               - & $46.5\pm8.6$ & $32.1\pm4.5$ & $15.2\pm3.0$ \\
%          $ 0.0$ & $28.9\pm 6.6$ & $26.3\pm4.5$ & $28.7\pm9.0$ & $17.8\pm3.7$ \\
%          $-0.2$ & $29.0\pm 3.1$ & $22.7\pm1.9$ & $18.7\pm4.1$ & $18.2\pm3.7$ \\
%    \end{tabular}
%    \caption{Saturated ion heat flux level and standard deviation to it for the different models considered and the three triangularity cases.}
%    \label{ch7_tab:multi_model_heat_fluxes}
%\end{table}

\section{Conclusions}
\label{sec:conclusions}

In this work, we apply the GM approach as a multi-fidelity tool to investigate the impact of triangularity on tokamak edge turbulence.
By using experimental data from a DIII-D discharge, we leverage the efficiency of the GM approach to conduct nonlinear GK simulations with realistic plasma edge geometry parameters.

A set of ten GMs, the \PJ{4}{2} basis, is sufficient to capture the essential features of TEM and ITG instabilities and turbulence.
This represents a significant and promising finding for the application of the GM approach in scenarios that also present kinetic electron instabilities.
We note that a six GMs system, based on the \PJ{2}{1} basis, also shows similar results but overestimates the transport and is more sensitive to the truncation closure \citep{Frei2023Moment-basedModel}.
The investigation of an improved closure scheme, which prevents this effect, is left for future work.

We identify the TEM stabilization as the main mechanism responsible for the improved confinement in NT configuration. 
When the TEMs are stabilized, the transport sensitivity to NT is reduced.
Consequently, one must expect a possible reduction of the benefit of NT in future tokamaks, with parameters such as ITER, where ITG-driven transport is expected.

The presence of the TEM instability at nominal parameters of the DIII-D discharge prevents simple electron models, such as the one of the AEM and RFM, from predicting accurate transport levels.
However, in the case of purely ITG-driven transport, the three models agree.
In fact, the AEM and RFM predict ion heat flux levels comparable to the simulations with kinetic electrons for $\delta = 0$ and $R_N=0$. 
The AEM captures the reduction of transport predicted in the KEM simulation at NT, with relatively small discrepancies rising from the destabilization of a TEM due to the increase of triangularity, in agreement with previous GK studies.
The RFM shows limited sensitivity to changes in the value of triangularity and agrees with the other models only when the density gradient is neglected. 
This is explained by comparing the GM amplitude between the AEM and RFM simulations, revealing an energy transfer toward higher-order GMs that cannot be captured in the RFM.

Finally, we explore the triangularity-gradient parameter space comparing scans obtained with KEM, AEM, and RFM simulations. 
The AEM simulations confirm previous findings that NT does not reduce the level of transport when far from marginal stability \citep{Merlo2023OnTokamaks,Balestri2024PhysicalPlasmas}. 
% However, considering the KEM simulations, the NT configuration is favorable when TEMs are present.
For the RFM simulations, we recover the confinement enhancement at NT for lower gradient strength, albeit with reduced magnitude compared to AEM results. 
For increased temperature gradient, the RFM predicts a confinement improvement for highly positive triangularity, contradicting both the KEM simulations and literature results.
Its limitations prove the importance of retaining kinetic effects to study the impact of triangularity on turbulence even in a purely ITG-dominated regime.
The KEM simulations demonstrate that the discharge \#186473 parameters lie close to a transition between ITG- and TEM-driven turbulence.
When the TEMs are stabilized, through a reduction of either triangularity or temperature and density gradient levels, the KEM simulations agree with the AEM results.
This indicates a minor role of the electrons when the TEMs are not present. 

The ability of the GM approach to retrieve ITG- and TEM-driven turbulence using only ten moments of the distribution function strongly reduces the costs of the presented simulation scans with respect to standard GK approaches.
It also suggests that a high-order RFM could capture the impact of triangularity in a broader parameter space.
The reduced numerical cost provides a valuable tool for studying a wide range of plasma turbulence phenomena and their intricate interactions.

\section*{Acknowledgements}
The authors acknowledge helpful discussions with J. Ball, S. Brunner, A. Balestri, A. Vol\v{c}okas, and R. Mackenbach.
The simulations presented herein were carried out in part on the CINECA Marconi supercomputer under the TSVVT422 project and in part
at CSCS (Swiss National Supercomputing Center). This work has been carried out within the framework of the EUROfusion Consortium, via the Euratom Research and Training Programme (Grant Agreement No 101052200 – EUROfusion) and funded by the Swiss State Secretariat for Education, Research and Innovation (SERI). Views and opinions expressed are, however, those of the author(s) only and do not necessarily reflect those of the European Union, the European Commission or SERI. Neither the European Union nor the European Commission nor SERI can be held responsible for them.

\appendix
\section{Electromagnetic nonlinear gyromoment hierarchy}
\label{app:emgmhierarchy}
We detail the terms present in the GM hierarchy in a flux tube configuration, Eq. \ref{eq:moment_hierarchy}.
The perpendicular magnetic term, related to the curvature and gradient drifts, writes
\begin{align}
    \mathcal M_{\perp a}^{pj} &= \frac{\tau_a}{q_a} \mathcal C_{k_x k_y} \squareparenthesis{\sqrt{(p+1)(p+2)} n_a^{p+2,j} + (2p+1)n_a^{pj} + \sqrt{p(p-1)}n_a^{p-2,j}}
    \nonumber\\&
    + \frac{\tau_a}{q_a} \mathcal C_{k_x k_y} \squareparenthesis{(2j+1)n_a^{pj} - (j+1)n_a^{p,j+1}-jn_a^{p,j-1}},
    \label{ch2_eq:Mperpapj}
\end{align}
with $C_{k_x k_y}$ the magnetic curvature operator, while the parallel magnetic term, related to the Landau damping and the mirror force, is expressed as
\begin{align}
    \mathcal M_{\parallel a}^{pj} =& 
    \frac{\hat B^{-1}}{J_{xyz}} \frac{\sqrt{\tau_a}}{\sigma_a}\left\{\ddz \aleph_a^{p\pm1,j}\right.
    - \ddz\ln B \left[(j+1)\aleph_a^{p\pm1,j}-j\aleph_a^{p\pm1,j-1}\right]
    \nonumber\\&
    \qquad\left. +\ddz\ln B\sqrt{p}\squareparenthesis{(2j+1)n_a^{p-1,j} -(j+1)n_a^{p-1,j+1} - jn_a^{p-1,j-1}}\right\}
    \label{ch2_eq:Mparapj}
\end{align}
with $\aleph_a^{p\pm1,j}=\sqrt{p+1} n_a^{p+1,j} + \sqrt{p} n_a^{p-1,j}$. 
The background gradient drift terms are
\begin{align}
    \mathcal D_{Na}^{pj} = \RNa\kernel_a^jik_y\Upsilon\delta_{p0},
    \label{ch2_eq:DNapj}
\end{align}
for the density and
\begin{align}
    \mathcal D_{Ta}^{pj} = \RTa \left\{\kernel_a^j\squareparenthesis{\frac{1}{\sqrt{2}}\delta_{p2} -\delta_{p0}}+ \squareparenthesis{(2j+1)\kernel_a^j-(j+1)\kernel_a^{j+1}-j\kernel_a^{j-1}}\delta_{p0}\right\}ik_y\Upsilon,
    \label{ch2_eq:DTapj}
\end{align}
for the temperature.
In Eqs. \ref{ch2_eq:Mperpapj}, \ref{ch2_eq:Mparapj} and \ref{ch2_eq:DTapj}, we introduce the non-adiabatic GM, 
$$
n_a^{pj}(\bm k,t)=N_a^{pj}+\frac{q_a}{\tau_a} \kernel_a^j\left(\phi\delta_{p0} - \frac{\sqrt{\tau_a}}{\sigma_a}A_{\parallel}\delta_{p1}\right).
$$
We also express the gyro-averaging operator using the Bessel function of the first kind, $J_0$, projected on the Laguerre basis, \begin{equation}
    J_0\left(\sqrt{\lperpa \wperpa}\right) = \sum_{n=0}^\infty \kernel_a^n(\lperpa)L_n(\wperpa).
\label{ch2_eq:bess_lag}
\end{equation}
with the kernel functions, $\kernel_a^n(\lperpa)=(\lperpa)^n e^{-\lperpa}/n!$, $\lperpa=\tau_a \sigma_a^2\kperp^2/2$ and $\kperp^2 = g^{xx}k_x^2 + 2 g^{xy}k_x k_y + g^{yy} k_y^2$ \citep{Frei2020}.\\
The nonlinear term related to the $\bm E\times \bm B$ drift is expressed as
\begin{align}
    \mathcal{S}_a^{pj} &=  \sum_{n=0}^{\infty}\left[\kernel_a^n\phi,\sum_{s=0}^{n+j}d_{njs} N_a^{ps}\right]_{k_x,k_y}\nonumber\\
    &- \sum_{n=0}^{\infty}\frac{\sqrt{\tau_a}}{\sigma_a}\left[\kernel_a^n A_\parallel,\sum_{s=0}^{n+j}d_{njs}\left(\sqrt{p+1} N_a^{p+1,s}+\sqrt{p}N_a^{p-1,s}\right)\right]_{k_x,k_y},
    \label{ch2_eq:sapj}
\end{align}
where $[f_1,f_2]_{k_x,k_y}$ denotes the evaluation of the Poisson bracket in Fourier space. 
In Eq. \ref{ch2_eq:sapj}, we use the Bessel-Laguerre decomposition, Eq. \ref{ch2_eq:bess_lag}, and we express the product of two Laguerre polynomials as a sum of single polynomials using the identity
\begin{equation}
L_jL_n=\sum_{s=0}^{n+j}d_{njs}L_s
\label{ch2_eq:lagprod}
\end{equation}
with
\begin{equation}
    d_{njs} = \sum_{n_1=0}^n\sum_{j_1=0}^j\sum_{s_1=0}^s \frac{(-1)^{n_1+j_1+s_1}}{n_1!j_1!s_1!}\binom{n}{n_1}\binom{j}{j_1}\binom{s}{s_1}.
    \label{ch2_eq:dnjs}
\end{equation}
Finally, we close our system with the dimensionless GK quasi-neutrality equation in Fourier space, i.e.
\begin{equation}
  \sum_a\left( \frac{q_a^2}{\tau_a}\squareparenthesis{1-\sum_{n=0}^{\infty}\left(\kernel^n_a\right)^2}\right)\phi
  =  \sum_a q_a\sum_{n=0}^{\infty}\kernel_a^n N_a^{0n},
    \label{ch2_eq:poisson_moments}
\end{equation}
and the GK Ampère equation
\begin{equation}
	\left(2\kperp^2 + \beta_e \sum_a \frac{q_a}{\sigma_a}\sum_{n=0}^\infty \left(\kernel^n_a\right)^2\right)A_\parallel 
	= \beta_e \sum_a q_a \frac{\sqrt{\tau_a}}{\sigma_a}\sum_{n=0}^\infty \kernel^n_a N_a^{1n}.
    \label{ch2_eq:Ampère_moments}
\end{equation}

% If we consider an adiabatic electron response, the GK quasi-neutrality equation Eq. \ref{ch2_eq:poisson_moments} writes
% \begin{equation}
%      \left( 1 + \frac{q_i^2}{\tau_i}\left[1-\sum_{n=0}^{\infty}\left(\kernel_{i}^n\right)^2\right]\right)\phi - \langle \phi \rangle_{yz} = q_i\sum_{n=0}^{\infty}\kernel_{i}^n N_i^{0n},
%     \label{ch5_eq:poisson_moments_adiabe}
% \end{equation}
% where $\langle \phi \rangle_{yz}$ is the flux surface average of $\phi$, namely
% \begin{equation}
%     \langle \phi \rangle_{yz} = \frac{1}{\int\dz J_{xyz}}\int\dz J_{xyz}\phi(k_x,k_y,z,t)\delta_{k_y0}.
% \end{equation}
% The flux surface average term results from the averaging of the asymptotically fast motion of the adiabatic electrons along the magnetic field line. 
% In this case, no electromagnetic fluctuation can be considered, which sets $A_\parallel  = 0$.
% \input{991_appendix_GM_GF_equivalency}

\bibliographystyle{jpp}
\bibliography{references}
\end{document}